\documentclass[]{revtex4}

\usepackage[usenames,dvipsnames]{color}
\usepackage{graphicx}
\usepackage{amsmath}
\usepackage{float}

\usepackage{ulem}

\begin{document}

\title{Overcoming damping in spin wave propagation: A continuous excitation approach to determine time-dependent dispersion diagrams in 2D magnonic crystals}

\author{B. Van de Wiele$^1$}
    \email{ben.vandewiele@ugent.be}
\author{F. Montoncello$^2$}

\affiliation{
  $^1$Department of Electrical Energy, Systems and Automation, Ghent University, Sint Pietersnieuwstraat 41, B-9000 Ghent, Belgium.\\
  $^2$Dipartimento di Fisica e Scienze della Terra, Universit\`a degli Studi di Ferrara, Italy.
}

\begin{abstract}
We propose an alternative micromagnetic approach to determine the spin wave dispersion relations in magnonic structures.  Characteristic of the method is that a limited area of the system is continuously excited with a spatially uniform oscillating field, tuned at a given frequency.  After a transitory time, the regime magnetization dynamics is collected and a spatial Fourier analysis on it determines the frequency vs wave vector relation.  Combining several simulations in any predetermined range of frequencies, at any resolution, we investigate the dispersion relations for different kinds of magnonic crystals: a dot array, an antidot array, and a bicomponent film.  Especially compared to traditional pulse-excitation methods this technique has many advantages.  First, the excitation power is concentrated at a single frequency, allowing the corresponding spin waves to propagate with very low attenuation, resulting in a higher k-space resolution. Second, the model allows to include very large wave vector components, necessary to describe the high-frequency response of non-quantized spin waves in quasi-continuous systems.  Finally, we address some possible experimental opportunities with respect to excitation/detection techniques over large distances and the observation of the odd/even symmetry of spin waves using Brillouin light scattering.
\end{abstract}

\maketitle

\section{Introduction}
Historically, research on patterned magnetic materials has first focused on dense arrays of non-interacting and hence independent dots, mainly for magnetic memory purposes. For this reason, the magnetic samples consisted often of arrays of disks and even rings, in the vortex state, or, in general, of systems with some anisotropy but vanishing stray fields \cite{Demokritov, Hubert, Novosad, Hillebrands, Bailleul, Hertel, Vogel, Schult, JAP2008}.  However, more recently the interaction among dots in an array is seen more as an opportunity rather than a limit, and the possibility to exploit the collective spin excitation (“magnons”) as information carriers has been valued, in analogy to light in photonic crystals \cite{Kruglyak, Lenk}. For this reason, interacting nano-magnets are called “magnonic crystals”.  They can be tailored to obtain magnonic band diagrams with the possible occurrence of magnonic bandgaps. The research includes any other magnetic “meta-material”, namely a material with periodic magnetic properties: antidot arrays \cite{Zivieri} and bicomponent continuous films \cite{Ma, Gubbiotti-2012} are other important examples. These systems are promising candidates for building novel versatile magnetic devices, which can operate either as waveguides or memories, but also as tunable filters, depending on the amplitude of an external magnetic field \cite{Demidov, Chumak, Chumak2}. Furthermore, the possibility of modifying the propagation properties of the information carriers is important for building magnonic- and spin-logic devices \cite{Ding, Lenk, Khitun}.

Besides the experimental investigation (fabrication and characterization of a magnonic crystal as well as the measurement of its dynamical properties), great effort has been devoted to the theoretical understanding of the properties of spin waves in such systems, in particular to set computational tools (either analytical or micromagnetic) suitable to simulate and predict their dynamic behavior \cite{Sukhostavets, yu, Puszkarski, Puszkarski2, Kraw, Kostylev, Verba, Montoncello-2012, Montoncello-2013, Giovannini, NMAG, OOMMF, Micromagus}.  Generally speaking, for a given equilibrium magnetic configuration, there are two main approaches in this context.  First, the simulation of an infinite, periodic 1-D or 2-D system by applying periodic boundary conditions on a unit cell. In the case of the dynamical matrix method \cite{Giovannini}, the Landau-Lifshitz equation is cast in a matrix form, which depends on the given Bloch wave vector.  The diagonalization of this matrix leads to frequencies and space profiles of all possible modes in the spectrum at the given Bloch wave vector. In the second approach a finite, though large, area of the magnetic system, excited by a pulse field with a given symmetry profile, is simulated.  After collecting the time evolution of the magnetic response throughout the sample the frequency and wave vector spectrum of the response is obtained through a temporal and spatial Fourier analyses \cite{Venkat, Krug}. Here, the excited mode type depends on the symmetry of the excitation pulse, while the minimum frequency and the frequency resolution depends on the pulse duration and simulation time window.  Comparing the two techniques, the latter has been widely preferred by experimentalists, especially for the possibility of controlling the input/output frequency bandwidth through the duration of the time pulse.

The approach we present in this paper belongs to the second category, but is innovative because the excitation field is continuously feeding the system at one single frequency.  Consequently, only the modes beating at that frequency are excited.  Contrary to a pulse excitation, only a spatial Fourier analysis is required to extract the wave vector spectra at the considered frequency.  By combining multiple simulations at different excitation frequencies one can determine the dispersion relations at the frequency range of interest at any frequency resolution.  Likewise, one can directly visualize the dynamic magnetization existing at a given frequency.  Moreover, by studying the dispersion graphs in the time domain, one can directly extract the temporal phase information of the magnonic waves with respect to the microwave excitation.  We will show that this point is crucial in understanding the real symmetry of the excited propagating mode.  Finally, the continuous excitation allows spin waves to propagate to a longer extent, overcoming damping problems typical for large arrays \cite{Vivas, NeusserDamp, KrugDamp}, and providing a larger resolution in wave vector space.

In the next section we illustrate the method, highlighting the differences with the classically adopted pulse excitation.  We then apply the method on 2D magnonic crystals of different kinds: a square array of interacting magnetic dots, a continuous film with antidot lattice, and a square bicomponent chessboard. In this way we show both the versatility of the method and validate it with (published) experimental data.

The purpose of this work is twofold: not only presenting a "new" way of finding mode dispersions with a higher resolution, but also a "proposal" for experimentalists to continuously feed the system with some excitation.  We will see that, as a consequence, classically hardly observable odd spin wave modes can in principle be detected by Brillouin light scattering.  Furthermore, the continuous excitation also helps to overcome damping criticality in magnonic crystals.  In this respect, we should remark that only in large magnonic crystals collective spin wave properties can be exploited.  Here, damping of the spin wave intensity is so critical that a signal, usually delivered by a pulse at the input side of the magnonic crystal, can hardly be detected at the output side, compromising the applicability of the technology.  Conversely, when reverting to smaller magnonic crystals to overcome the complete signal damping, the collective propagation properties are suppressed by stationary modes of the lattice as a whole.  Therefore, in the context of magnonic- and spin-logic devices, a continuous excitation of the spin waves is a promising alternative to address both the collective properties of spin modes and low signal damping.

\section{Methodology}

\subsection{General excitation}

Let us consider a general field $H^{exc}(t)$ which excites spin waves in a magnetic system.  The excited spin waves will depend on the frequency spectrum present in the excitation
\begin{equation}
\begin{split}
 H^{exc}(t) &= \int C^{exc}(f) e^{\imath 2\pi f t}\,\mathrm{d}f\\
  &= \int \left\{A^{exc}(f) + \imath B^{exc}(f)\right\}e^{\imath 2\pi f t}\,\mathrm{d}f\\
  &=\int |C^{exc}(f)|\cos\{2\pi f t + \phi^{exc}(f)\}\,\mathrm{d}f \\
  &\quad + \imath\int |C^{exc}(f)|\sin\{2\pi f t + \phi^{exc}(f)\}\,\mathrm{d}f\\
  &=\int |C^{exc}(f)| e^{\imath[2\pi f t + \phi^{exc}(f)]}\,\mathrm{d}f.
\end{split}\label{FFT_exc}
\end{equation}
Indeed, a general excitation contains a distribution of frequencies with corresponding amplitudes 
\begin{equation}
 |C^{exc}(f)| = \sqrt{\left[A^{exc}(f)\right]^2 + \left[B^{exc}(f)\right]^2} \label{ampl_exc}
\end{equation}
and corresponding phases 
\begin{equation}
 \phi^{exc}(f)=\arctan\frac{A^{exc}(f)}{B^{exc}(f)}. \label{psi_exc}
\end{equation}
In terms of phasors rotating in frequency space, $|C(f)|$ is the amplitude of the phasor, while $\phi^{exc}(f)$ is its initial phase angle.

Due to the excitation, the magnetization $\mathbf{M}(\mathbf{r}, t)$ varies in time and space.  We now study the spin waves propagating along the $x$-axis by analyzing the $z$-component of the magnetization along this axis.  This is done by extracting the spin wave modes defined by a characteristic wave vector $k_x$ and frequency $f$ by means of a spatial and a temporal Fourier transform of $M_z(x,t)$ respectively.
\begin{equation}
\begin{split}
M_z&(x,t) = \iint C^{M_z}(k_x,f) e^{\imath 2\pi f t} e^{\imath 2\pi k_x x}\,\mathrm{d}f\,\mathrm{d}k_x\\
  &= \iint \left\{A^{M_z}(k_x,f) + \imath B^{M_z}(k_x,f)\right\}\\
  &\qquad\qquad\qquad\qquad \times e^{\imath 2\pi (ft+k_xx)}\,\mathrm{d}f\,\mathrm{d}k_x\\
  & = \iint |C^{M_z}(k_x,f)| \,\times \\
&e^{\imath \left[ 2\pi(ft+k_xx) + \phi^{exc}(f) + \phi^{M_z}(k_x,f)\right]}\,\mathrm{d}f\mathrm{d}k_x.\label{FFT_Mz}
\end{split}
\end{equation}
Expression (\ref{FFT_Mz}) describes the dispersion relation of the different spin wave modes with given ($k_x,f$). The amplitude 
\begin{equation}
|C^{M_z}(k_x,f)| = \sqrt{\left[A^{M_z}(k_x,f)\right]^2 + \left[B^{M_z}(k_x,f)\right]^2},\label{ampl_Mz}
\end{equation}
depends on the amplitude of the excitation field at that frequency $|C^{exc}(f)|$ as well as on the geometrical and material properties of the sample supporting this mode.  Hence, since one is only interested in the latter, one should correct expression (\ref{ampl_Mz}) for the excitation field amplitude (\ref{ampl_exc}).  

The phase has different contributions
\begin{equation}
\begin{split}
 \phi^{exc}(f) + \phi^{M_z}&(k_x,f)\\&=\arctan\frac{A^{M_z}(k_x,f)}{B^{M_z}(k_x,f)} \label{psi_Mz}.
\end{split}
\end{equation}
It indeed depends on the phase of the considered frequency component in the excitation $\phi^{exc}(f)$.  Moreover it contains an additional phase difference $\phi^{M_z}(k_x, f)$.  The latter can be interpreted as a temporal phase difference between the mode and the excitation ---a phase $\phi^{M_z}(k_x,f)=0$ then stands for a mode beating in phase with the excitation--- or as a spatial phase difference between spin wave modes.  Due to the propagating character of the spin waves, both interpretations are equivalent.

\subsection{Pulse excitation}

In micromagnetic simulations, a Gaussian pulse profile $e^{-at^2}$ ($a$ real, positive) is often used to excite the spin modes.  This pulse has a spectrum
\begin{equation}
C^{exc}(f) = \sqrt{\frac{\pi}{a}}e^{\pi^2f^2/a},
\end{equation}
yielding a constant phase angle $\phi^{exc}(f) = 0$.  In most cases, the pulse is applied on a restricted area of the sample.  After applying the pulse, the magnetization dynamics is simulated for some time $T_{sim}$, sampling the magnetization data at $N_t$ time instants. This leads to a frequency resolution $\Delta f = 1/T_{sim}$
\begin{equation}
 f_n = \frac{n}{T_{sim}} \quad n=0\ldots N_t/2-1. \label{frequencies}
\end{equation}
Only $N_t/2$ frequency components are obtained due to the cyclic nature of Fourier transforms.  Expression (\ref{frequencies}) shows that, aiming at a large frequency resolution and a large frequency range, one needs to simulate the magnetization response for a long simulation time $T_{sim}$ and store the magnetization multiple times $N_t$.    Based on one single simulation one can build the complete dispersion diagram describing the amplitude of the different excited modes $|C^{M_z}(f_n,k_{x,m})|$ with 
\begin{equation}
 k_{x,m} = \frac{m}{L_{x,sim}} \quad m=0\ldots N_x/2-1 \label{wave_vectors}
\end{equation}
the discretized wave vector.  Here, $L_{x,sim}$ is the sample length in the studied $x$-direction, discretized using $N_x$ discretization cells.  Equivalently, to obtain a large $k$-space resolution over a large $k$-range, one needs to simulate a long sample $L_{x,sim}$ discretized with a large amount of discretization cells.  

Hence, if one is aiming for a large frequency and $k$-space resolution, the magnetization processes in very large samples are to be computed during very long time windows $T_{sim}$.  However, while initially only the magnetization at the excited area is affected by the field pulse, the propagating spin waves distribute the mode energy through the complete sample leading to a decreasing signal amplitude with time, even when a zero Gilbert damping is considered.  This puts an upper limit on useful simulation windows $T_{sim}$ and sample dimensions $L_{x,sim}$ and thus intrinsically limits the frequency and wave vector resolution.  This a fortiori applies for the real experiment, where the propagating signal strength decreases even more because of Gilbert damping.

\subsection{Sinusoidal excitation}
Let us now consider a continuous sinusoidal excitation, applied on some restricted area of sample
\begin{equation}
 H^{exc}(f_0,t) = h \sin(2\pi f_0 t). \label{Hexc_sine}
\end{equation}
Spin waves with the same frequency $f=f_0$, but different $k_x$ wave vector, can now propagate from the excited area along the studied propagation direction.  To determine these modes one only needs to perform a Fourier analysis in $k$-space
\begin{equation}
\begin{split}
 M_z&(x,f_0,t) = \int C^{M_z}(k_x,f_0,t)e^{\imath 2\pi k_x x}\, \mathrm{d}k_x\\
& =\int \left\{A^{M_z}(k_x,f_0,t) + \imath B^{M_z}(k_x,f_0,t)\right\}\\
& \qquad \qquad \times e^{\imath 2\pi k_x x }\, \mathrm{d}k_x\\
& =\int |C^{M_z}(k_x,f_0,t)|e^{\imath \left[ 2\pi k_x x + \phi^{M_z}(k_x, f_0)\right]}\, \mathrm{d}k_x\\
& =\int |C^{M_z}(k_x, f_0)|\sin\{2\pi f_0 t + \phi^{M_z}(k_x,f_0)\}\\
& \qquad \qquad \times e^{\imath 2\pi k_x x }\, \mathrm{d}k_x.
\end{split}\label{FFT_Mz_sine}
\end{equation}
Indeed, the mode varies sinusoidally in time with the same frequency as the excitation and a possible phase delay $\phi^{M_z}(k_x, f_0)$.  
Expression (\ref{FFT_Mz_sine}) shows that the mode amplitude depends on the time instant $t$ at which the magnetization is sampled.  In a simulation, the excitation is first applied for a given number of periods $T_0=1/f_0$ to ensure that the excited modes can propagate into the structure and no transient effects are present.  After this, the magnetization is sampled at $N_t$ time instants during half an excitation period
\begin{equation}
 t_i=\frac{i}{N_tf_0}\quad i=0\ldots N_t/2-1 . \label{time_samples}
\end{equation}
By inspecting at what time instant $t_{max}=i_{max}/N_tf_0$ the mode amplitude $|C^{M_z}(k_x,f_0,t)|$
is maximal, one can extract the mode phase angle with respect to the excitation as
\begin{equation}
 \phi^{M_z}(k_x,f_0) = \pm\frac{\pi}{2} - \frac{i_{max}}{N_t}\pi \label{phase_angle}
\end{equation}
and the effective mode amplitude as
\begin{equation}
|C^{M_z}(k_x, f_0)| = |C^{M_z}(k_x,f_0,t_{max})|.
\end{equation}
Note that $\phi^{M_z}(k_x,f_0)$ is determined up to an angle $\pi$.  This explains why only half a period is sampled: the other half results in identical amplitudes $|C^{M_z}(k_x, f_0)|$ and opposite phases $\phi^{M_z}(k_x,f_0)$.  A visual inspection of the magnetization distribution at $t=t_{max}$ easily defines the sign  $\pm\pi$.  Furthermore, we have chosen the excitation to follow a sine and not a cosine function to limit transient effects when starting the simulations.  Indeed, starting from $t=0$ the excitation increases gradually from zero towards its maximum value.

To build the complete dispersion diagram, one needs to perform multiple simulations with excitation frequencies distributed over a predefined range.  This distribution does not need to be uniform: certain frequency ranges can be simulated with an increased frequency resolution, e.g. to study a specific spin wave mode or a band gap in large detail, while other frequency ranges can be sampled with a lower frequency resolution by considering fewer simulations,  which makes the method extremely versatile.

Contrary to applying a pulse excitation, it is not necessary to consider a vanishing damping in order to ensure a spin wave propagation far into the sample.  The external sinusoidal field feeds the system continuously leading to a large amplitude wave propagation over extended areas of the magnonic crystal.  Consequently, in regime, the Gilbert damping balances the energy that is continuously introduced into the system by the excitation as one would find in the real experiment.  Moreover, the total power is not dissipated among all the modes over a wide range of frequencies as with a pulse excitation, but concentrated at the frequency of a single mode (or just a few, if more k-modes occur at that frequency).   Hence the continuous excitation approach opens opportunities  with respect to the effective implementation of spin wave based technology where large distances should be covered.

Some other straightforward advantages of the method: (i) By considering a fixed excitation amplitude $h$ in all simulations for different frequencies $f_0$, see expression (\ref{Hexc_sine}), all modes are excited with the same power, making---contrary to an pulse excitation---a direct comparison between all mode amplitudes possible; (ii) The magnetization profile and its time variation at a given frequency is directly obtained from the simulations enabling an immediate visual interpretation of the possible modes existing at that frequency; (iii) For the complete dispersion spectrum, the phase difference $\phi^{M_z}(k_x,f)$ with the excitation can be obtained in a straightforward manner.

In this approach, many small simulations instead of one single large simulation (pulse excitation) have to be performed.  This is in line with current computer hardware evolutions towards powerful computer clusters.  All simulations for different frequencies $f_0$ are completely independent from each other and can thus be performed in parallel. For this contribution we use our GPU cluster containing 16 nodes.  Each node in the cluster runs the micromagnetic software package MuMax, speeding up each single simulation with two orders of magnitude \cite{VAN-11}.

\section{Dispersion diagram of a 2D magnonic crystal}
We apply the sinusoidal excitation approach to the 2D magnonic crystal presented by Tacchi \textit{et al.} \cite{TAC-10}.  It comprises a 30\,nm thick matrix consisting of 450\,nm$\times$450\,nm square Permalloy dots separated 70\,nm from each other, hence with lattice constant U=520\,nm.  We use identical magnetic parameters: saturation magnetization $M_s$=820\,kA/m, exchange constant $A$=1.0$\times10^{-11}$\,J/m and zero anisotropy. An external bias field of 119.4\,kA/m is applied parallel to one of the dot edges. The propagation properties of the waves traveling perpendicular to and parallel with this bias field are studied. 

First, we recall that in the linear approximation of magnetization dynamics, the time evolution of the magnetization can be seen as
\begin{equation}
 \mathbf{M}(\mathbf{r},t)=\mathbf{M}(\mathbf{r},0)+\delta\mathbf{m}(\mathbf{r},t),
\end{equation}
where $\mathbf{M}(\mathbf{r},0)$ is the equilibrium 2D map of the magnetization, and $\delta\mathbf{m}(\mathbf{r},t)$ is the dynamic magnetization which we will study in the following.  

In order to understand the collective spin wave dynamics in arrays of dots at any given time, dynamic modes are expressed as 2D Bloch waves
\begin{equation}
 \delta\mathbf{m}(\mathbf{r}) = \delta \tilde{\mathbf{m}}_{\mathbf{k}}(\mathbf{r})e^{\imath\mathbf{k}\cdot\mathbf{r}}, \label{Bloch}
\end{equation}
with $\mathbf{r}$ the radius-vector in direct space, $\mathbf{k}$ the Bloch wave vector, and $\delta\tilde{\mathbf{m}}_{\mathbf{k}}$ the cell function, which contains the periodicity of the array. At $\mathbf{k}$=(0,0) the mode is represented at any $\mathbf{r}$ by its cell function.  This cell function can be described as for single dots, by means of the orientation of the nodal lines (surfaces) with respect to the (local) direction of the magnetization. Consequently, Damon-Eshbach-like (n-DE) modes are characterized by $n$ nodal lines
parallel to the local magnetization while Backward-like (m-BA) modes are characterized by $m$ nodal lines perpendicular to the local magnetization.  The fundamental mode can be seen either as 0-BA or 0-DE and mixed modes possess nodal lines of both types.  Furthermore, confinement effects give rise to end-modes, with dynamic magnetization intensity localized only at the edges of the dot in the direction of the applied field.  These modes usually appear at the lowest frequencies \cite{jorzick, JAP2008}.  Due to the Bloch factor, the apparent profile $\delta\mathbf{m}(\mathbf{r})$ of all these modes changes when $\mathbf{k}\neq\mathbf{0}$ at different $\mathbf{r}$ in the array.   In our simulation approach, one can directly visualize the 2D profile $M_z(x,y,f_0)$ of each mode at any frequency and label it.  By means of a spatial Fourier analysis of the magnetization patterns in a given frequency range, the full spectrum of spin waves can be determined.  Their behavior --propagating or stationary-- can be deduced by inspecting the slope of the dispersion curves.

\begin{figure}
\begin{center}
\includegraphics[width=0.45\columnwidth]{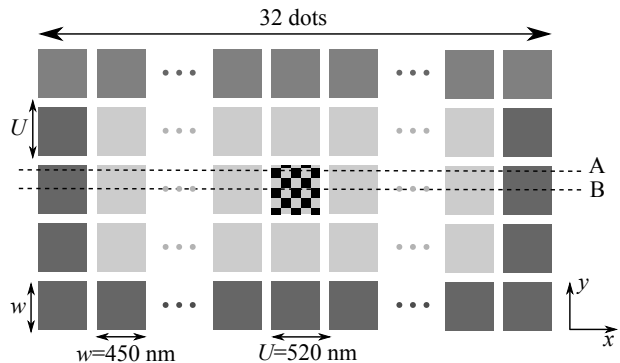}
\end{center}
\caption{The simulated geometry is a Permalloy dot array containing 32$\times$5 dots.  The outer dots (dark gray) have a higher damping constant to inhibit unwanted reflections.  The excitation is applied on a central dot depicted with a chessboard pattern.  The spin wave data is obtained along horizontal lines in the central array of dots, e.g. line 'A' for edge modes and line 'B' for bulk modes.   \label{Fig_geometry}}
\end{figure}

\begin{figure*}
\begin{center}
\includegraphics[width=\textwidth]{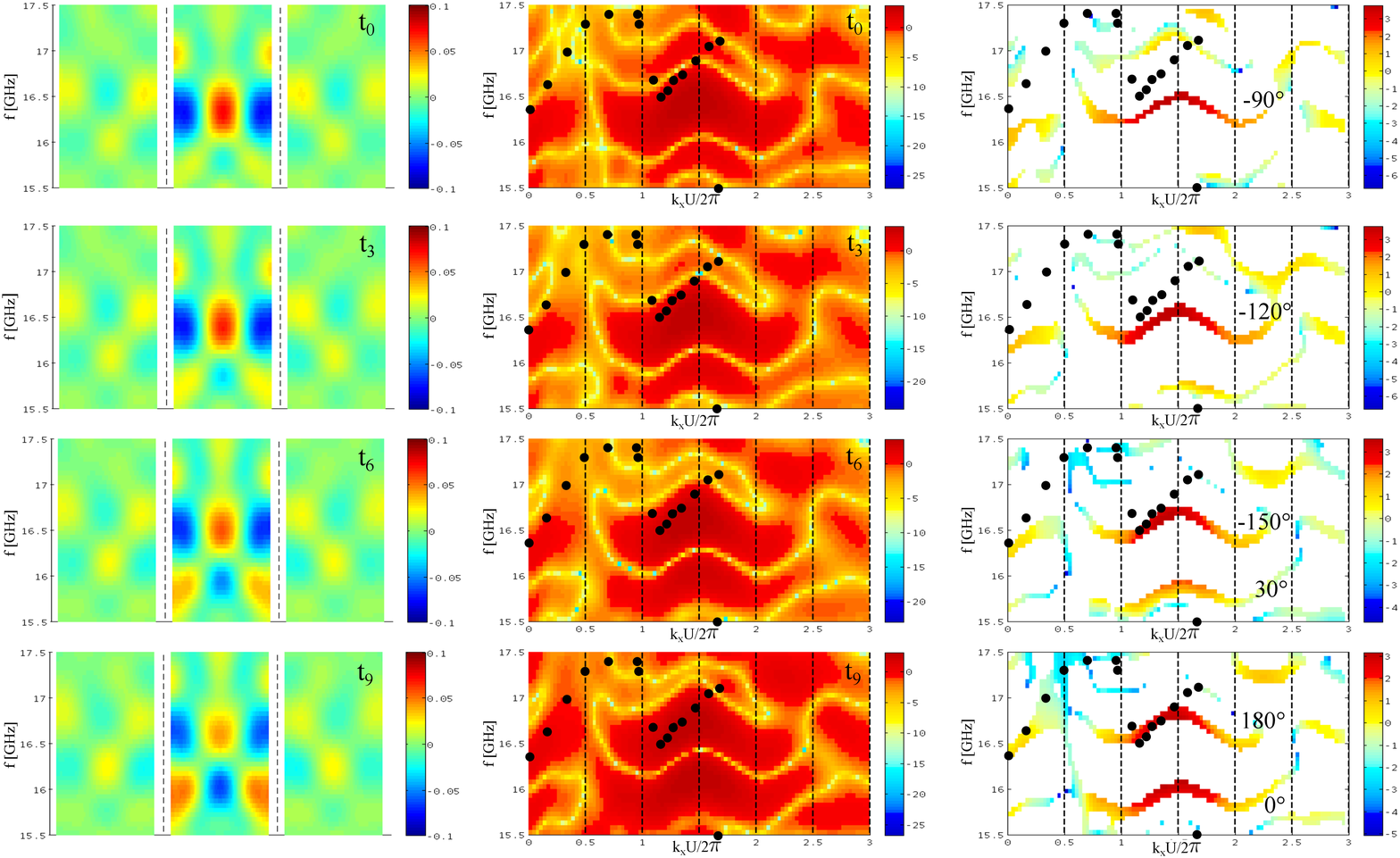}
\end{center}
\caption{Left column: magnetization profile $M_z(x,f_0,t_i)$ for different time points $t_i$ depicted in the top-right corner of every panel. Each horizontal line is a magnetization profile for a given frequency $f_0$ along a line 'B' in Fig. \ref{Fig_geometry}, here limited to three central unit cells.  Middle column: corresponding dispersion diagram of the mode amplitudes $|C^{M_z}(k_x,f_0,t_i)|$ obtained after a spatial Fourier transform of $M_z(x,f_0,t_i)$.  Right column: dispersions of modes with maximum power at the considered time point $|C^{M_z}(k_x,f_0)| = |C^{M_z}(k_x,f_0,t_{max})|$ and the corresponding phase difference with respect to the excitation. The full circles are the experimental BLS data from \cite{TAC-10}. \label{Fig_phases}}
\end{figure*}

\subsection{Simulation procedure}
We simulate the dot array by considering an elongated rectangular matrix of 32$\times$5 dots, see Fig. \ref{Fig_geometry}, discretized using 4096$\times$512 finite difference cells.  Only spin waves propagating along the $x$-axis are studied.  Hence, the spectra for perpendicular and parallel bias field are obtained from distinct simulations with the bias field applied along the $y$-axis and the $x$-axis, respectively.  The long simulation size $L_{x,sim}=16.64\,\mu$m ensures a high $k$-space resolution, see (\ref{wave_vectors}).  Magnetization data $M_z(x,t)$ can be collected along horizontal lines in the central row of the array, e.g. line 'A' for edge modes and line 'B' for bulk modes.  We treat five rows of dots as a good approximation of the full 2D character of the magnonic crystal.  The dots at the boundary of the crystal---in Fig. \ref{Fig_geometry} depicted in dark gray---have a high damping constant.  In this way, spin waves are not reflected, but absorbed without interfering with the spin waves in the central row of dots. This rules out stationary modes due to confinement effects.  Elsewhere, the damping constant is put at $10^{-4}$.   As in \cite{TAC-10}, a spatially uniform excitation profile applied on a central dot, in Fig. \ref{Fig_geometry} depicted with the chessboard pattern is used.  However, the excitation field is not a pulse, but has a continuous sinusoidal wave form [eq. (\ref{Hexc_sine})].  The amplitude $h$ is two orders of magnitude smaller than the bias field.  This is larger than in the experiment, but is required to have a spin wave signal which is above the noise level throughout large portions of the sample.  For our GPU computations this relative noise level is $\mathcal{O}(10^{-6})$ since we have a floating point accuracy.  The amplitude is however small enough in order not to awake non-linear magnetization dynamics.  For each direction of the bias field we have run simulations with excitation frequencies $f_0$ ranging from 5\,GHz to 20\,GHz with steps $\Delta f_0$=50\,MHz.

\subsection{Phase sensitive dispersion diagrams}

\begin{figure}
\begin{center}
\includegraphics[width=0.5\columnwidth]{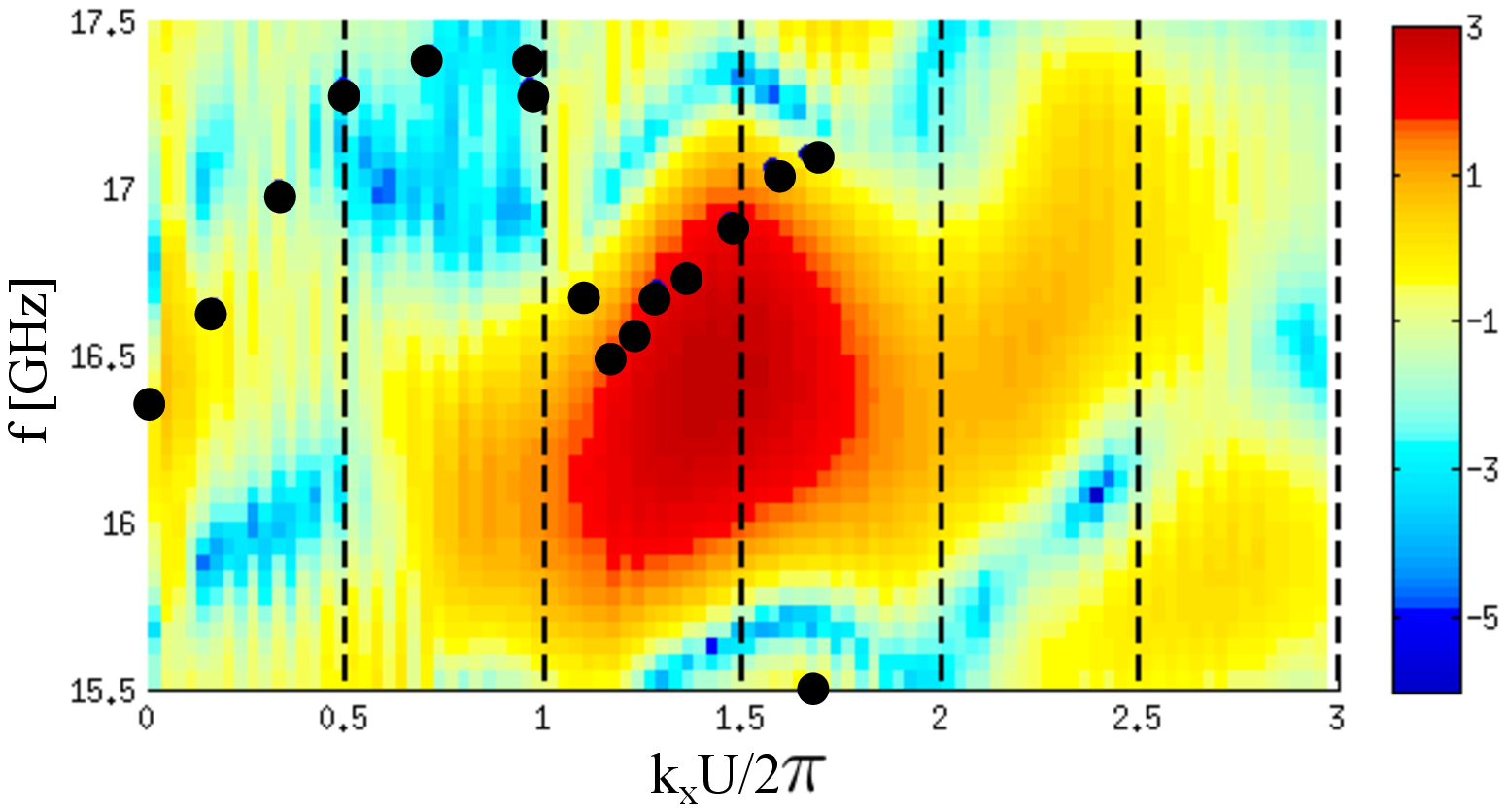}
\end{center}
\caption{Mode amplitude $|C^{M_z}(k_x,f_0)|$ constructed by summing up all phase sensitive phase diagrams.  The full circles are the experimental BLS data from \cite{TAC-10}.\label{Fig_amplitude}}
\end{figure}

To explain the phase extraction method, we consider now bulk spin waves propagating perpendicular to the bias field with 15.5\,GHz$<f_0<$17.5\,GHz.  A discussion of the complete dispersion diagram is postponed to the next Section.  Once the system is relaxed to its ground state with the bias field directed along the $y$-axis, the sinusoidal excitation is applied for 20 periods to enable the spin waves to propagate in the system and reach equilibrium conditions between excitation and damping.  During the 21$^{th}$ excitation period, $M_z(x)$ along line 'B' in Fig. \ref{Fig_geometry} is stored at time points
\begin{equation}
 t_i = \frac{i}{36} \frac{1}{f_0}\qquad i=0\ldots17.
\end{equation}
In Fig. \ref{Fig_phases}, left column, each panel shows the 1D spin wave profile $M_z(x,f_0,t_i)$ along line 'B' in Fig. \ref{Fig_geometry} for successive frequencies $f_0$ at a different time point $t_i$.  While the magnetization profile is only shown in three central dots, the corresponding dispersion diagram $|C^{M_z}(k_x,f_0,t_i)|$, presented in the second column of Fig. \ref{Fig_phases}, is obtained by means of a spatial Fourier transform on the complete data set $M_z(x=0\rightarrow L_{x,sim}, f_0, t_i)$.  
In the third column of Fig. \ref{Fig_phases} we plot only ($k_x$,$f_0$) points where the intensity in the power spectrum is maximal at time points $t_i$.  From that, the phase difference with the excitation $\phi^{M_z}(k_x, f_0)$ can be found using (\ref{phase_angle}), resulting in phase sensitive dispersion diagrams.  Combining all maximum values in one graph, i.e. collecting all mode amplitude data irrespective the phase, one can visualize the mode amplitude  $|C^{M_z}(k_x,f_0)|$ as in Fig. \ref{Fig_amplitude}.

\begin{figure*}
\begin{center}
\includegraphics[width=0.5\columnwidth]{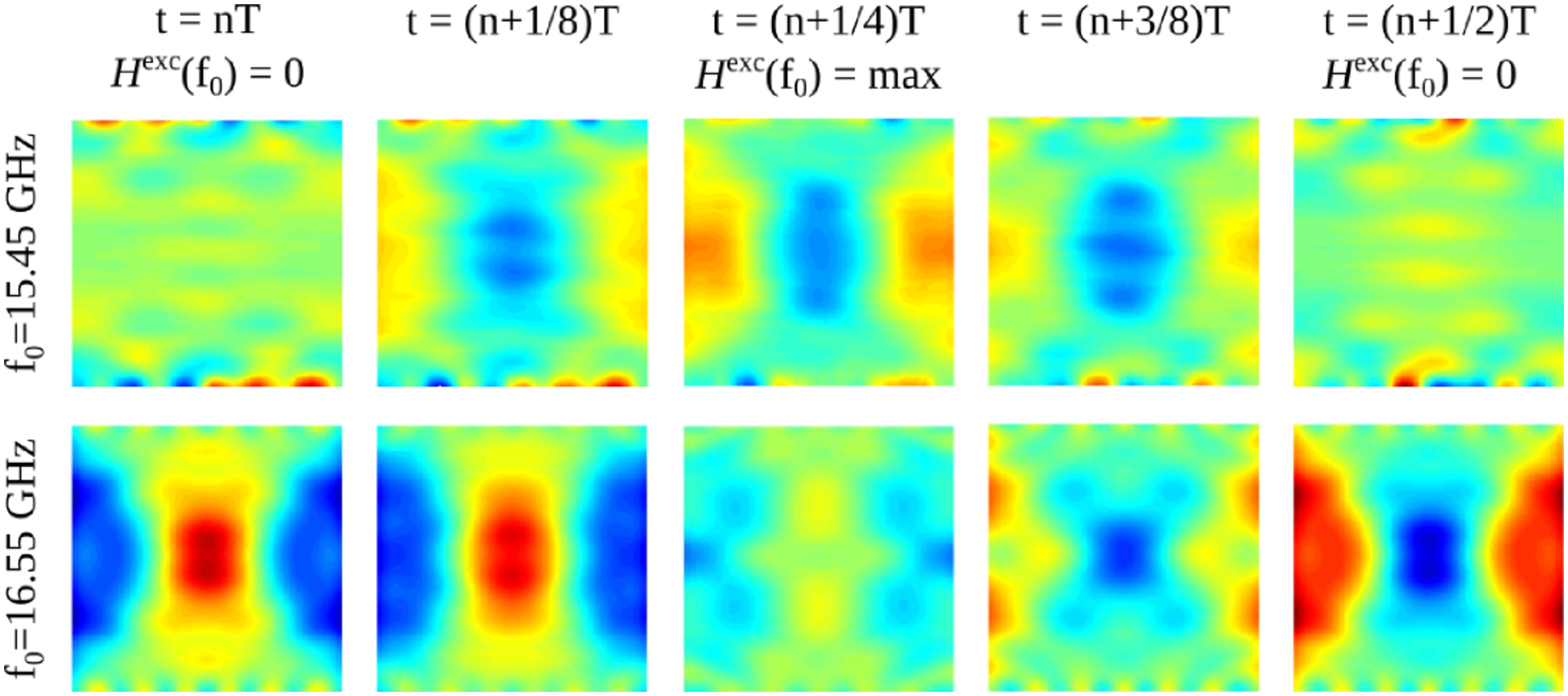}\hspace{0.5cm}
\includegraphics[width=0.5\columnwidth]{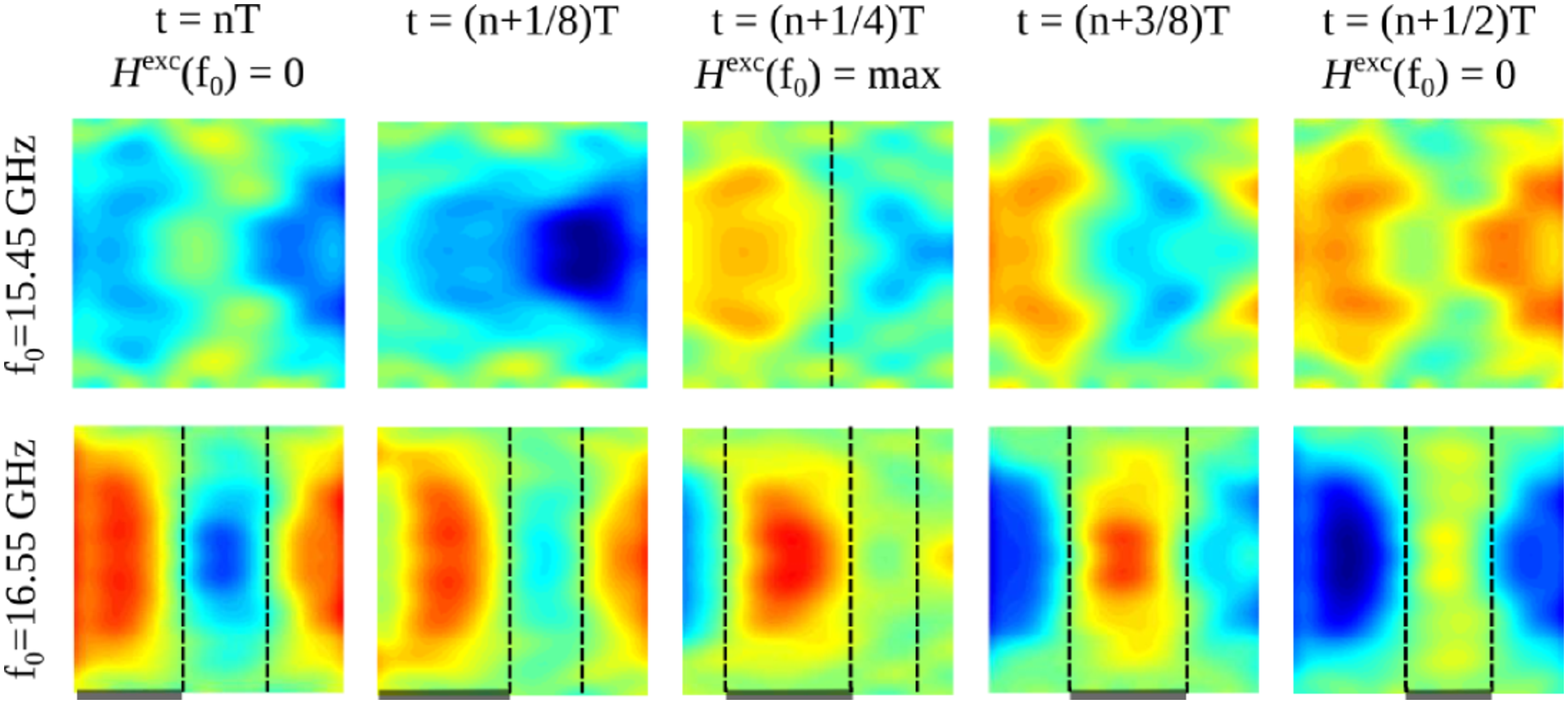}
\end{center}
\caption{Evolution of the magnetization patterns at successive time points (0, 1/8, 1/4, 3/8, 1/2 of the period T) in the excited, central cell of the array (left panel) and in the neighboring cell on the right of the excited one (right panel) obtained at a frequency $f_0$=15.45\,GHz and $f_0$=16.55\,GHz.  Blue/red/green colors correspond to a local negative/positive/zero out-of-plane magnetization.  The gray lines underneath the bottom-right series of magnetization patterns emphasize the propagating character of the spin wave mode. \label{Fig_primitive}}
\end{figure*}

From a closer inspection of the graphs in Fig. \ref{Fig_phases}-left column, and Fig. \ref{Fig_amplitude}, we see that --depending on the time phase with respect to the excitation-- a 2-DE-like profile is established in the excited, central cell over a 2\,GHz frequency range.  Indeed, at different time points, the modes in this frequency range give rise to the same 2-DE-like profile.  This is illustrated by the maximum power lines which, remarkably, shift from one frequency to another as time flows indicating that the phase difference $\phi^{M_z}(k_x,f)$ varies continuously with frequency, see the phase dependent dispersions in Fig. \ref{Fig_phases}-right column.   Figure \ref{Fig_primitive}-left shows dynamic magnetization maps of the central, excited cell at different time points.   For $f$=15.45\,GHz [$f$=16.55\,GHz] this dot has a maximal [zero] magnetization amplitude when the excitation is maximal and a zero [maximal] magnetization amplitude when the excitation is zero, meaning that the magnetization profile is beating in phase [with a 90 degree phase difference] with respect to the excitation.

However, the dynamic magnetization map in the central dot is ambiguous, since it only visualizes how the spin waves are excited by the sinusoidal excitation field, but not how they propagate through the magnonic crystal.  In fact, since the considered excitation has a uniform profile covering the entire central cell, the magnetization map in this specific cell should possess an even symmetry.  Indeed, it can be either uniform or have an even number of nodes in both the $x$- and $y$-direction (as even-node harmonics of higher order).  This should be true at all time points. Consequently, the spin waves need to be stationary: their nodal lines can not move.  We can understand this by considering that, in the excited cell, the stationary wave results from a superposition of two propagating waves $\psi_{-\bf{k}}$ and $\psi_{+\bf{k}}$ with exactly the ‎same cell function, but traveling in opposite directions out of the central cell: $-\bf{k}$ and +$\bf{k}$.  The superposition can lead to either functions $\psi_a ‎‎=\psi_{-\bf{k}}+\psi_{+\bf{k}}$ or $\psi_b =\psi_{-\bf{k}}-\psi_{+\bf{k}}$.   To have an even profile, the function $\psi_a$ needs to ‎be a superposition of even functions $\psi_{-\bf{k}} (\psi_{+\bf{k}})$, with a resulting power that is maximum at t=0, and zero at t=T/4, i.e. has a 90 degree phase difference with respect to the excitation (see Fig. \ref{Fig_primitive}-left, mode at 16.55 GHz).   For the same reason, $\psi_b$ needs to be a superposition of odd functions with a resulting amplitude that is exactly zero at t=0, ‎and maximum at t=T/4, i.e. it beats in phase with respect to the excitation (see Fig. \ref{Fig_primitive}-left, mode at 15.45 GHz).  

Hence, looking at the excited cell only, it is impossible to study the propagating character of the single $\psi_{\pm\bf{k}}$ functions: one needs to consider a unit cell that is not involved in the excitation process.  To that purpose, Fig. \ref{Fig_primitive}-right shows the time evolution of the dynamic ‎magnetization in the first cell on the right of the excited one, the $\psi_{+\bf{k}}$ wave.  Here, the propagating character is observed by following the shift of extrema and nodal lines from the left to the right.  For example, referring to the mode at 16.55\,GHz, it is clear that the maximum on the left at t=0 (the red area) gradually shifts to the right part of the dot as time increases.  Similar considerations hold for all other modes at different frequencies.

Moreover, in Fig. \ref{Fig_primitive}-left, we can see that both modes, at 15.45\,GHz and 16.55\,GHz, show an even profile in the excited cell.  However, by inspection of the profiles on the right panel, we understand that the propagating mode $\psi_{+\bf{k}}$ at 16.55\,GHz is actually a 2-DE mode (even) and that one at 15.45\,GHz is actually a 1-DE mode (odd).  Following the explanations above, the superposition of left and right propagating waves results, in the excited cell (left panel), to a maximal signal at t=0 for the even mode at 16.55\,GHz and a maximal signal at t=T/4 for the odd mode at 15.45\,GHz.
The fact that a continuous, spatially uniform excitation can activate modes independent of their symmetry by just tuning the excitation frequency is a remarkable feature of the presented method.
 
Going through the broad band shown in Fig. \ref{Fig_amplitude}, at each frequency f a different combination of propagating modes $\psi_{f,+\bf{k}}$ and $\psi_{f,-\bf{k}}$ superimposes to the same magnetization profile in the excited cell: the 2-DE profile.  In fact, the magnetization profile can result simply from a single mode (odd or even), or, more often, can result from the superposition of several modes as in the case of non-monotonic dispersion curves $\omega(k)$.  The time point at which the superposition gives rise to the maximum signal defines the phase difference with the excitation.  This phase difference turns a magnetization profile with any definite symmetry into an even profile in the central excited cell as required by the excitation process.


\subsection{Complete dispersion diagrams}

\begin{figure*}
\begin{center}
\includegraphics[width=\textwidth]{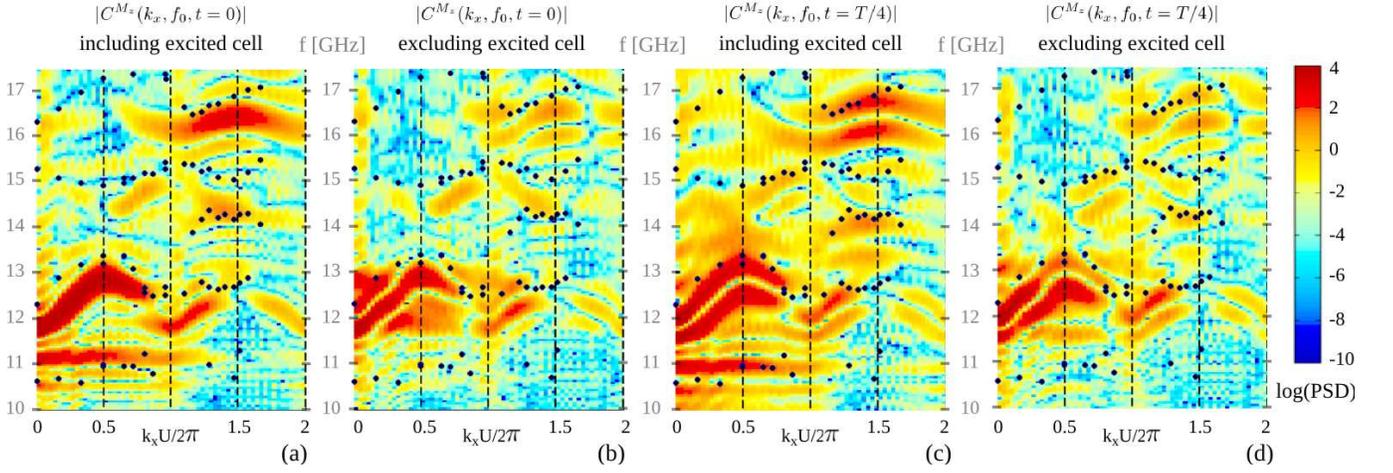}
\end{center}
\caption{Dispersion relations in the Voigt geometry ($\mathbf{k}\perp\mathbf{H}$) at time points t=0 (panel a,b) and t=T/4 (panel c,d), obtained from a spatial Fourier analysis including (panel a,c) or excluding (panel b,d) the central cell. Full circles are the BLS measurements taken from ref. \cite{TAC-10}. \label{Fig_TAC}}
\end{figure*}
 
\begin{figure*}
\begin{center}
\includegraphics[width=\textwidth]{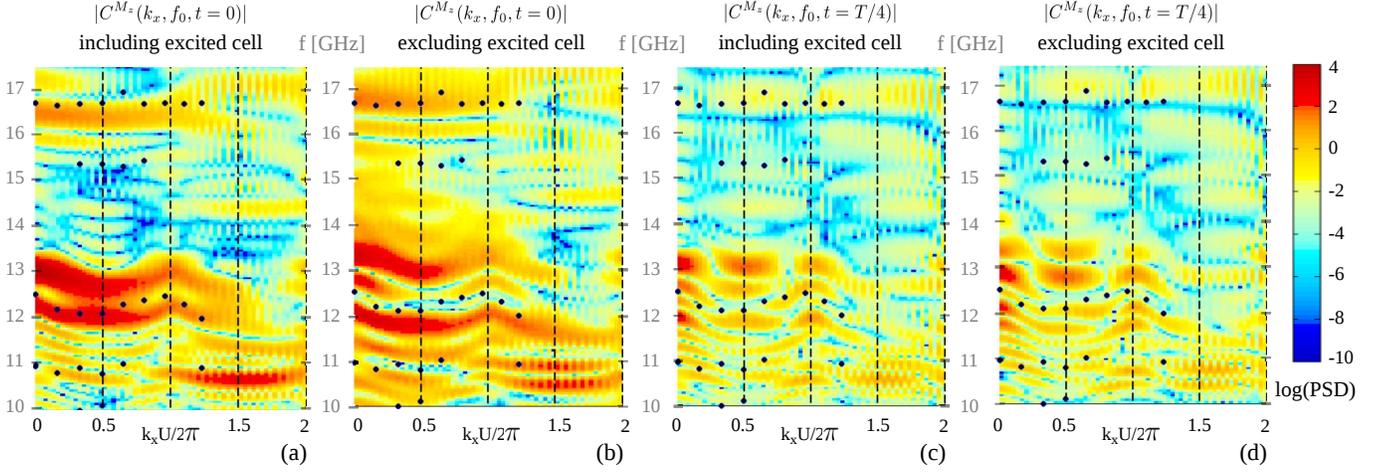}
\end{center}
\caption{Dispersion relations in the backward geometry ($\mathbf{k}\parallel\mathbf{H}$) at time points t=0 (panel a,b) and t=T/4 (panel c,d), obtained from a spatial Fourier analysis including (panel a,c) or excluding (panel b,d) the central cell.  Full circles are the BLS measurements taken from ref. \cite{TAC-10}. \label{Fig_TAC_par}}
\end{figure*}

We have explained why and how the mode power depends on time, and how this dependence is different if we focus on the central dot (the excited one, where spin waves have a stationary behavior) or in the rest of the array (where the propagating character of waves becomes apparent). In particular, when performing the spatial Fourier analysis, we can take or not into consideration the central (excited) cell, and understand its influence on the dispersion graphs.  

In an ideal infinite array, in which edge effects are negligible and no direct excitation is considered, only thermal spin waves exist.  Even though their profile varies in time and from cell to cell, these variations are undetectable when averaged over the crystal, leading to a calculated/measured power independent of time.  On the contrary, when in this infinite array the waves are excited from a given lattice site, the mode power averaged over the array is weighted by the intensity of the wave at different lattice sites.   Consequently, now the mode power definitely depends on time as the Gilbert damping makes an infinite array effectively finite.  Indeed, modes with $\mathbf{k}\neq0$ have different dynamic profiles in adjacent cells of the crystal, see eq. (\ref{Bloch}).  As time flows, the wave propagates and damps out, shifting the different magnetization profiles from one cell to another and decreasing their contribution to the corresponding calculated/measured average power.
  
This actually happens in our simulations, which deal both with a finite size array and with a localized excitation field, especially when we exclude the central, excited cell from the Fourier analysis. In Fig. \ref{Fig_TAC} [Fig. \ref{Fig_TAC_par}] we plot the dispersion relations of the spin waves propagating perpendicular [parallel] to the applied field at time points t=0 and t=T/4, calculated including (panel a,c) or excluding (panel b,d) the excited cell in the center of the array.  While the dispersion graphs that exclude the central cell show a small time dependence, only due to the propagation and damping of modes described above, when the central cell is included in the Fourier analysis the time dependence is more critical.  Due to the excitation process and superposition effects, modes in the central cell have a power which strongly varies with time: namely, modes with even [odd] profile are particularly amplified at t=0 [t=T/4].   Recalling that the BLS cross section is, at a first approximation, proportional to the average mode power \cite{Gub2005}, this leads to the proposal to measure the presence of odd modes.

Usually, in a BLS experiment, odd modes have a very small cross section, making them hardly measurable, unless considering large incidence angles \cite{Gub2005}.  However, in an excitation/detection experiment involving the BLS experiment, the laser beam (which, in microfocused-BLS-technique \cite{MicroBLS}, has a space FWHM of about 300 nm) can be focused to illuminate an area containing or not the excited region, and the spectra collected in the two cases can be compared. The peaks corresponding to odd modes will be present only in the spectrum containing the excited cell while those corresponding to even modes will be present in both spectra.  In this way, the spin wave dynamics of magnonic systems can be interpreted with much more confidence.

When comparing our results with the experiments, we recall that the usual resolution of BLS measurements is around 0.5\,GHz. This suits with the experimental points found for k=0 at about 10.6\,GHz and 11.0\,GHz in the first panels of both Figs. \ref{Fig_TAC} and \ref{Fig_TAC_par}, which should indeed correspond to the same mode.  To this extent, the simulations show a good agreement with BLS measurements, especially for t=0, where even (BLS active) modes are amplified. 

Referring to Figs. \ref{Fig_TAC} and \ref{Fig_TAC_par}, we discuss now only the few modes that are helpful to interpret the experimental BLS points.  The mode at about 11\,GHz (at k=0) is rather non-dispersive along the direction perpendicular to the bias field (Fig. \ref{Fig_TAC}-panel a, group velocity nearly zero), ‎and has a slightly negative dispersion along the direction parallel to the bias field (Fig. \ref{Fig_TAC_par}-panel a); from inspection of the magnetization profile, we find that its cell function is rather uniform ‎parallel with the bias field, but with 4 nodes along perpendicular direction.  This mode is usually addressed to ‎as a 4-BA-like mode, and propagates parallel to the direction of the applied field. Since this is a ‎mode with an even cell function, it is best seen in panel (a) of both Figs. \ref{Fig_TAC}, \ref{Fig_TAC_par}. We skip modes in the range 11$\div12$\,GHz, which show a clear backward-like character, but are outside the scope of this paper.

We focus now our attention on the mode that experimentally occurs at about 12.3\,GHz (k=0, mean value between the two BLS points at $\mathbf{k}\parallel\mathbf{H}$ and $\mathbf{k}\perp\mathbf{H}$), and which we find numerically at about 12.1\,GHz.  This is the fundamental mode, which has a rather uniform profile, the largest bandwidth ($\omega_{X(Y)}-\omega_{\Gamma}$) in both X and Y directions ‎and largest intensity in the first Brillouin zone (BZ, 0$\div$0.5, in units of 2$\pi$/U).
This mode stems from the Damon-Eshbach (DE) surface mode of ‎the continuous film.  However, because of Bragg diffraction at the void spacers between adjacent dots, its dispersion is ‎discontinuous at zone boundaries and band gaps arise.   With reference to Fig. \ref{Fig_TAC}, in the second BZ ‎‎(k$\in$ 0.5$\div$ 1.0$\times$ 2$\pi$/U, and $\omega/2\pi\in14.75\div$15.25 GHz), due to the Bloch factor [eq.(\ref{Bloch})], the DE mode gets a profile with one nodal ‎line, we call it 1-DE, and in the $2^{nd}$ BZ it gets the largest intensity in the spectrum.  As can be expected from an odd mode, the corresponding dispersion, folded in the $1^{st}$ BZ (in reduced scheme), has a vanishing intensity, if the central dot is not ‎considered (Fig. \ref{Fig_TAC}), or an appreciable one if the central dot is considered (at k=0 it occurs at about 15.2\,GHz). \\
The DE mode in the range (1.0$\div$1.5, in units of 2$\pi$/U, in the frequency range 16.25$\div$17 GHz) has 2 ‎nodal lines, and is called the 2-DE mode when folded to the $1^{st}$ BZ (at k=0 it occurs at about 16.3\,GHz).  In the range (1.5$\div$2.0, $1^{st}$ half of ‎the $3^{rd}$ BZ) the DE mode gets 3 nodes, and when considering it in the $1^{st}$ BZ (where it shows vanishing power) it is usually addressed as 3-DE mode (and so on for increasing Bloch wavevector). The same mode, considered along the direction parallel to the applied field (Fig. \ref{Fig_TAC_par}), has a negative dispersion, because it stems from the backward mode of the continuous film, while the 1-DE and 2-DE modes discussed above show in that zone only poor or slightly negative dispersion.  The behavior of the dispersion curves is found as predicted from the effective vector model, introduced in ref. \cite{TacPRL2011}: the effective wave vector, which is limited to the reduced zone, corresponds to the wave vector of the dipolar spin waves in an effective continuous film that replaces the magnonic crystal.  This representation enables an easy understanding of the dispersion properties for modes with complex profiles appearing in non-trivial lattice symmetries \cite{Montoncello-2013}.\\
 
\section{Dispersion diagrams in an antidot array and in a bicomponent magnonic crystal}
Referring to the paper of Zivieri \textit{et al}. \cite{Zivieri} we apply the presented method to a square lattice of holes with 120 nm diameter, with lattice constant $U$=800\,nm, in a 22\,nm thick permalloy film. Since the primitive cell of this system consists of a square with a central hole, it can be seen as the reverse image of a dot, i.e. as an \textit{antidot}. We take the same magnetic parameters used in that paper, namely saturation magnetization $M_s$=748\,kA/m and exchange stiffness parameter $A$=1.3$\times 10^{-11}$\,J/m. Discretization cells of 3.125$\times$3.125$\times$22 nm$^3$ are used and the magnetic bias field $H$=15.9\,kA/m is applied parallel to one primitive vector of the lattice.

\begin{figure}
\begin{center}
\includegraphics[width=0.45\columnwidth]{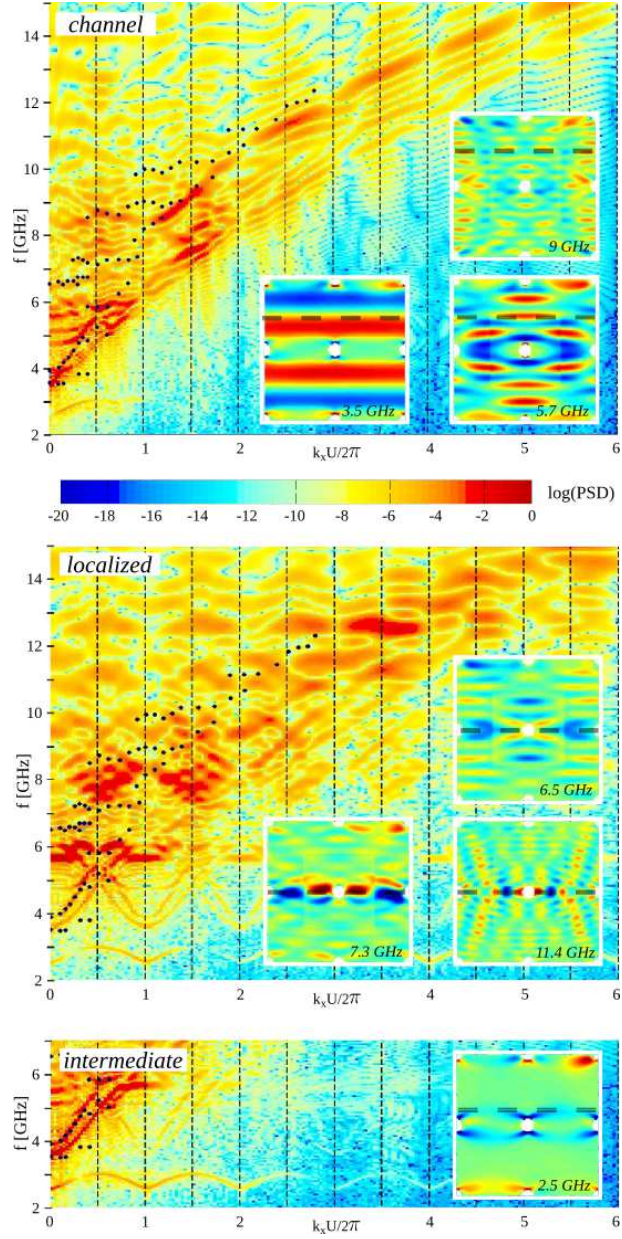}
\end{center}
\caption{Dispersion relations in the Voigt geometry ($\mathbf{k}\perp\mathbf{H}$) of spin waves in an antidot array. Full circles are the BLS measurements taken from ref. \cite{Zivieri}. The top panel, labeled \textit{channel}, shows the dispersions of modes extending in the region between the holes, indicated by the dashed lines on the corresponding mode profiles, shown as insets.  The central panel, labeled \textit{localized}, shows dispersion of modes localized between adjacent holes along the propagation direction, as indicated by the dashed line in the insets. The panel at the bottom, named \textit{intermediate}, shows dispersion for modes with intermediate localization degree, as shown by the profile in the inset, and by the dashed line. \label{Fig_ZIV}}
\end{figure}

In Fig. \ref{Fig_ZIV} we plot the dispersion relations for modes propagating perpendicular to the applied field.  Depending on the line along which the magnetization data is collected and analyzed, we probe different modes, with a different localization of their dynamic magnetization:  halfway between the holes [through the holes], we probe the so-called \textit{channel modes} [\textit{localized modes}] (top [middle] panel).  The bottom panel is obtained considering the magnetization data along an intermediate line.  In the insets, the collection region is highlighted by dashed lines.  Depending on the collection region, the dispersions change because different modes, with a maximum intensity in different areas of the array, contribute to the graphs. 

As found in ref. \cite{Zivieri}, in this system the modes with the largest power (and BLS cross section) are the \textit{channel modes} that have their largest intensity situated between the succeeding columns of holes. These modes stem from the DE surface mode: when in the first BZ, the mode is called \textit{fundamental}, here found at about 3.8\,GHz (not shown). A higher order mode of the same family (1-DE mode), but occurring at zone boundary at 9\,GHz, is shown in the upper panel.  These modes have nodal lines that are (roughly) parallel to the applied field with mode propagation perpendicular to it. 
Along the channel we can also find BA-like modes, stemming from the backward volume mode of the continuous film.  Their nodal lines are  perpendicular to the direction of the applied field: for instance, the mode at 3.5\,GHz shown in the inset of Fig. \ref{Fig_ZIV}-\textit{channel} is a 1-BA mode at zone boundary, that one at 5.7\,GHz is a 4-BA (though rather wiggled).  

We now consider \textit{localized} modes with largest intensity along a line through the holes, perpendicular to the applied field as shown in the insets of Fig. \ref{Fig_ZIV}-\textit{localized}.   A few examples of their profiles are given in the insets: they are characterized by DE-like nodes (parallel to the applied field), markedly one for the mode at 7.3\,GHz, five for the mode at 11.4\,GHz. The mode at 6.5\,GHz is rather uniform and intense in the region between the holes, but with slight undulations outside that region.
In some cases, both dispersion maps highlight the same modes (e.g. the modes at 5.7\,GHz and 6.5\,GHz) indicating that the mode has a large amplitude in both regions of the antidot array.

Finally, there are modes localized in an \textit{intermediate} way: in Fig. \ref{Fig_ZIV}-\textit{intermediate} we show, as an example, the mode at 2.5\,GHz, which is a localized one (as apparent from the space profile in the inset) and has a definite propagating behavior, and a definite coherence up to many BZs. In ref. \cite{Zivieri} this mode was only mentioned as EM (edge mode) but no specific profile was shown.

\begin{figure}
\begin{center}
\includegraphics[width=0.45\columnwidth]{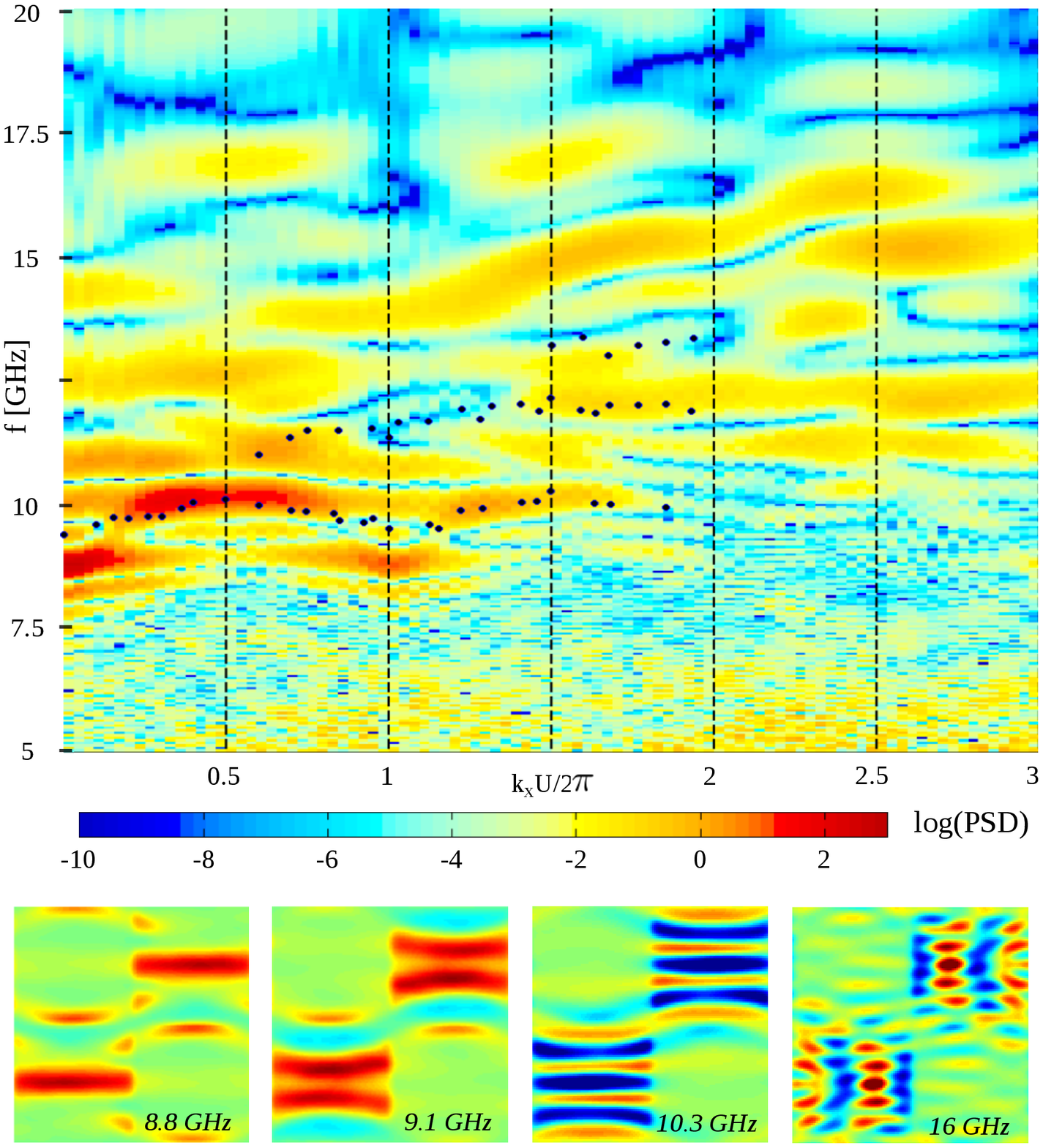}
\end{center}
\caption{Dispersion relations in the Voigt geometry ($\mathbf{k}\perp\mathbf{H}$) of spin waves in the bicomponent magnonic crystal. Full circles are the BLS measurements taken from ref. \cite{Gubbiotti-2012}. The panels at the bottom show four representative modes at different Bloch wave vector and frequency.  \label{Fig_GUB}}
\end{figure}

To give an idea of the broad applicability of the presented approach, we apply it also to a bicomponent system, namely a continuous 35\,nm thick film with a chessboard pattern of alternating 400$\times$400 nm$^2$ squares of permalloy and cobalt.  We refer to the paper of Gubbiotti \textit{et al.} \cite{Gubbiotti-2012}, and use the same geometry and magnetic parameters. For permalloy [cobalt], the saturation magnetization is $M_s$=760 [1250] kA/m, and the exchange stiffness is $A$=1.3 [3.0] $\times 10^{-11}$\,J/m. 

In the top panel of Fig. \ref{Fig_GUB}, we show the dispersion curves for modes propagating perpendicular to the applied field $H=79.5$\,kA/m.  The largest power is once more corresponding to the DE-like mode, which is recognizable at least up to the 3rd BZ, with increasingly smaller gaps (due to Bragg diffraction among different materials). In the bottom panel of the same figure, we show the magnetization maps of some spin modes aiming at a sound interpretation of the mode dynamics: at 8.8\,GHz we find the mode with the largest power and a rather uniform profile, usually addressed to as the fundamental mode; at 9.1\,GHz we show a backward-like mode with two nodal lines (2-BA mode) and k=0, which however happens to hybridize with the fundamental mode as k increases and eventually, at zone boundary, gets the largest power (i.e., represents the DE surface mode at $k_xU/2\pi=0.5$). To give an idea of other modes, we show also the backward-like  6-BA mode (at 10.3\,GHz and k=0) and the the 3-DE mode at 16\,GHz and $k_xU/2\pi=0.5$ (which corresponds to the DE surface mode at the end of the 3rd BZ, $k_xU/2\pi=2.5$).  We recover the observation made in the cited paper, namely that, at low frequencies, the spin wave amplitude is larger in the permalloy regions of the primitive cell.

As a final remark, concerning the excitation/detection of a spin-wave-information-carrier, we observe how in this system (as well as in antidot array) excited waves have a larger chance to arrive at the edges of the magnonic crystal, compared with other patterned systems with much more void spaces (e.g. the dot array).

\section{Conclusions}

We have presented a method to compute the dispersion diagrams of spin waves and get information on spin wave space profiles.  Unlike existing methods, this one deals with the continuous application of a sinusoidal field with a constant frequency on a restrictred area of system.  Consequently, only spin waves with the same frequency are excited.  A spatial Fourier analysis of the magnetization data makes it possible to study the dispersion relations $\omega(k)$ in the time domain: the dispersion graphs are time dependent, as they are determined by the phase difference of the mode with respect to the excitation. 
Dispersion relations in any arbitrary frequency range are obtained by combining independent simulations at different frequencies allowing any frequency resolution and making the approach perfectly suitable for a computational cluster environment, where many parallel simulations can be performed at once.

Since the excitation is continuous and spatially uniform, the activated spin waves are --in the excited area-- stationary with an even profile.  We show that these profiles can result from the superposition of propagating modes with either even or odd cell functions.  Consequently, the superposition of odd modes, usually have a vanishing average power anywhere in the crystal, give rise to an even profile with large average power in the excited area.   In this respect, we indicate that it can be experimentally feasible to both excite and detect the odd modes, and distinguish them from the even modes by comparison of BLS spectra measured in two different regions of the crystal, one containing the excitation area, the other not.  Odd spin wave profiles result in BLS peaks only visible in the region containing the excited area, while even spin wave profiles are visible in both regions.

Furthermore, due to the persistence of the excitation at one single frequency, the spin waves can propagate over very long distances through the crystal without severe attenuation.  This results in dispersions with both a large resolution in k-space, and a large intensity up to high order Brillouin zones. The latter is particularly interesting.  When dealing with almost continuous media (as antidot arrays or bicomponent materials), modes with quasi-continuous dispersions are possible.  For these modes it is necessary to look at large wave vectors if the higher frequency response of the mode is aimed at.  Moreover, the relatively low attenuation of the spin wave signal opens perspectives to technological applications in which larger systems with input and output antennas at opposite ends can be used as magnonic guides.

\section*{Acknowledgements}
We thank the group of Surface Magnetism of the University of Perugia (http://ghost.fisica.unipg.it/) for sharing with us the experimental data and useful discussions.  We thank A. Vansteenkiste for the numerical support. BVdW is financially supported by the Flanders Research Foundation (FWO). FM is financially supported by the Italian MIUR-PRIN 2010-11 Project 2010ECA8P3 \textit{DyNanoMag}.

\end{document}